\documentclass[11pt]{article}
\usepackage{amsmath}
\usepackage{graphicx}
\usepackage{amsfonts}
\usepackage{amssymb}
\usepackage{stmaryrd}
\usepackage{amscd}
\newcommand{\bc}{\begin{center}}
\newcommand{\ec}{\end{center}}
\newcommand{\be}{\begin{equation}}
\newcommand{\ee}{\end{equation}}
\newcommand{\bea}{\begin{eqnarray}}
\newcommand{\eea}{\end{eqnarray}}
\newcommand{\ba}{\begin{array}}
\newcommand{\ea}{\end{array}}
\newcommand{\edc}{\end{document}}

\def\O{\Omega}

\def\s{\sigma}

\begin{document}
\thispagestyle{empty}
\begin{center}

{\bf ON PERIODIC GIBBS MEASURES FOR POTTS-SOS MODEL ON THE CAYLEY TREE}\\
\vspace{0.4cm}  M.A.Rasulova \footnote{\it Namangan State University, Uychi str. 316, Namangan,
 Uzbekistan, e-mail: m$\_$rasulova$\_$a@rambler.ru}\\

\vspace{0.5cm}

{\bf Abstract}

\end{center}

In this paper we give a description of periodic Gibbs measures for
Potts-SOS model on the Cayley tree of order $k\geq 1$, i.e. a
characterization of such measures with respect to any normal
subgroup of finite index of the group representation of a Cayley
tree.



\begin{center}
\textbf{Introduction}
\end{center}

One of the central problems in the theory of Gibbs measures (GMs) is to describe
infinite-volume (or limiting) GMs corresponding to a given Hamiltonian. The existence of such measures for a wide class of
Hamiltonians was established in the ground-breaking work of Dobrushin (see, e.g., Ref. 1). However, a complete analysis of
the set of limiting GMs for a specific Hamiltonian is often a difficult problem.

In this paper we consider models with a nearest neighbour interaction on a
Cayley tree (CT). Models on a CT were discussed in Refs. 2 and 4-6. A classical
example of such a model is the Ising model, with two values of spin,1. It was
considered in Refs. 2, 6, 14, 16, 17 and became a focus of active research in the first half
of the 90s and afterwards; see Refs. 7-14.

In [18] all translation-invariant Gibbs measures for the Potts
model on the CT are described. In [19], [20] periodic Gibbs
measures, in [21] weakly periodic Gibbs measures for the Potts
model are studied.

In [22], [23] translation-invariant and periodic Gibbs measures
for the SOS model on the CT are studied.

Model considered in this paper (Potts-SOS model) are
generalizations of the Potts and SOS (solid-on-solid) models. In
[15] translation-invariant Gibbs measures for the Potts-SOS model
on the CT are studied, but periodic Gibbs measures were not
studied yet. In this paper we will study periodic Gibbs measures
for this model.

\begin{center}
\textbf{Preliminaries and main facts}
\end{center}
\textbf{The Cayley tree}. The Cayley tree $\Gamma^k$ (See [14]) of
order $ k\geq 1 $ is an infinite tree, i.e., a graph without
cycles, from each vertex of which exactly $ k+1 $ edges issue. Let
$\Gamma^k=(V, L, i)$ , where $V$ is the set of vertices of $
\Gamma^k$, $L$ is the set of edges of $ \Gamma^k$ and $i$ is the
incidence function associating each edge $l\in L$ with its
endpoints $x,y\in V$. If $i(l)=\{x,y\}$, then $x$ and $y$ are
called {\it nearest neighboring vertices}, and we write $l=<x,y>$.

 The distance $d(x,y), x,y\in V$ on the Cayley tree is defined by the formula
$$
d(x,y)=\min\{ d | \exists x=x_0,x_1,...,x_{d-1},x_d=y\in V \ \
\mbox{such that}  \ \
 <x_0,x_1>,...,<x_{d-1},x_d> \}.$$

For the fixed $x^0\in V$ we set $ W_n=\{x\in V\ \ |\ \
d(x,x^0)=n\},$
$$ V_n=\{x\in V\ \ | \ \  d(x,x^0)\leq n\}, \ \
L_n=\{l=<x,y>\in L \ \ |\ \  x,y\in V_n\}. \eqno (1) $$
 Denote $|x|=d(x,x^0)$, $x\in V$.

A collection of the pairs $<x,x_1>,...,<x_{d-1},y>$ is called a
{\sl path} from $x$ to  $y$ and we write $\pi(x,y)$ .
 We write $x<y$ if
the path from $x^0$ to $y$ goes through $x$.

It is known (see [14]) that there exists a one-to-one
correspondence between the set  $V$ of vertices  of the Cayley
tree of order $k\geq 1$ and the group $G_{k}$ of the free products
of $k+1$ cyclic  groups $\{e, a_i\}$, $i=1,...,k+1$ of the second
order (i.e. $a^2_i=e$, $a^{-1}_i=a_i$) with generators $a_1,
a_2,..., a_{k+1}$.

Denote $S(x)$ the set of "direct successors" of $x\in G_k$. Let
$S_1(x)$ be denotes the set of all nearest neighboring vertices of
$x\in G_k,$ i.e. $S_1(x)=\{y\in G_k: <x,y>\}$ and $ x_{\downarrow}
=S_1(x)\setminus S(x)$.

\textbf{ The model and a system vector-valued functional
equations.} Here we shall give main definitions and facts about
the model. Consider models where the spin takes values in the set
$\Phi=\{0,1,2,...,m\}, m\geq 1$. For $A\subseteq V$ a spin {\it
configuration} $\s_A$ on $A$ is defined as a function
 $x\in A\to\s_A(x)\in\Phi$; the set of all configurations coincides with
$\Omega_A=\Phi^{A}$. Denote $\O=\O_V$ and $\s=\s_V.$  Define a
{\it periodic configuration} as a configuration $\s\in \O$ which
is invariant under a subgroup of shifts $K\subset G_k$ of finite
index. More precisely, a configuration $\s \in \Omega $ is called
$K$ -periodic if $\s (yx)=\s (x)$ for any $x\in G_k$ and $y\in K$.

  For a given periodic configuration  the index of the subgroup is
called the {\it period of the configuration}. A configuration
 that is invariant with respect to all
shifts is called {\it translational-invariant}. Let
$G_k/K=\{K_0,K_1...,K_{r-1}\}$  factor group, where $K$ is a
normal subgroup of index $r\geq 1$.

The Hamiltonian  of the Potts-SOS model with nearest-neighbor
interaction has the form
$$
H(\s)=-J \sum_{<x,y>\in L}|\s(x)-\s(y)|- J_p \sum_{<x,y>\in
L}\delta_{\s(x)\s(y)}   \eqno (2)
$$
where $J, J_p\in R$ are nonzero coupling constants.

Let $h:x\mapsto h_{x}=(h_{0,x},h_{1,x},...,h_{m,x})\in R^{m+1}$ be
a real vector-valued function of $x\in V \backslash \{x^{0}\}$.
Let take into consideration the probability distributions
$\mu^{n}$ on $\Phi^{V_{n}}$ for given $n=1,2,...$ defined by
$$
\mu^{(n)}(\sigma_{n})=Z_{n}^{-1}\exp\big(-\beta
H(\sigma_{n})+\sum_{x\in W_{n}}h_{\sigma(x),x}\big)  \eqno (3)
$$
where $\sigma_{n}\in\Phi^{V_{n}}$ and the related partition
function $Z_{n}$ can be expressed as follows:
$$
Z_{n}(\widetilde{\sigma}_{n})=\sum_{\widetilde{\sigma}_{n}\in
\Phi^{V_{n}}}\exp\big(-\beta H(\widetilde{\sigma_{n}})+\sum_{x\in
W_{n}}h_{\widetilde{\sigma}(x),x}\big).  \eqno (4)
$$
We say that the probability distributions $\mu^{(n)}$ are
compatible if $\forall n\geq1$ and
$\sigma_{n-1}\in\Phi^{V_{n-1}}$:
$$
\sum_{\omega_{n}\in\Phi^{V_n}}\mu^{(n)}\big(\sigma_{n-1}\vee\omega_{n}\big)=\mu^{(n-1)}\big(\sigma_{n-1}\big),
\eqno (5)
$$
here $\big(\sigma_{n-1}\vee\omega_{n}\big)\in\Phi^{V_n}$ is the
concatenation of $\sigma_{n-1}$ and $\omega_{n}$.

\textbf{Definition 1.} If probability distribution $\mu^{(n)}$ on
$\Phi^{V_{n}}$ holds the equality $(5)$, we can say that a unique
measures $\mu$ on $\Phi^{V}$ exists such that, $\forall n$ and
$\sigma_{n}\in\Phi^{V_{n}},
\mu\big({\sigma|V_{n}=\sigma_{n}}\big)=\mu^{(n)}(\sigma_{n})$. The
measure $\mu$ is called splitting Gibbs measure
$\big($\textbf{SGM}$\big)$ corresponding to Hamiltonian $H$ and
function $h:x\mapsto h_{x}, x\neq x^{0}$.

The next theorem expresses the requirement on function $h$ that
the distributions $\mu^{(n)}(\sigma_{n})$ hold the compatibility
conditions.

\textbf{Theorem 1.}[15] Defined as in (3) the probability
distributions $\mu^{(n)}(\sigma_{n}), n=1,2,...,$ satisfy the
compatibility condition if and only if
$$
h_{x}^{*}=\sum_{y\in S(x)}F\big(h_{y}^{*},m,\theta,r\big)  \eqno
(6)
$$
is satisfied for any $x\in V \backslash \{x^{0}\}$, where
$$
\theta=\exp(J\beta),   \ \ r=\exp(J_{p}\beta)  \eqno (7)
$$
and also $\beta=1/T$ is the inverse temperature. Here $h_{x}^{*}$
represents the vector
$(h_{0,x}-h_{m,x},h_{1,x}-h_{m,x},...,h_{m-1,x}-h_{m,x})$ and the
vector function $F(.,m,\theta,r):R^{m}\rightarrow R^{m}$ is
defined as follows
$$
F(h,m,\theta,r)=\big(F_{0}(h,m,\theta,r),F_{1}(h,m,\theta,r),...,F_{m-1}(h,m,\theta,r)\big)
$$
where
$$
F_{i}(h,m,\theta,r)=\ln\frac{\sum_{j=0}^{m-1}\theta^{|i-j|}r^{\delta_{ij}}e^{h_{j}}+\theta^{m-i}r^{\delta_{mi}}}{\sum_{j=0}^{m-1}\theta^{m-j}r^{\delta_{mj}}e^{h_{j}}+r},
\eqno(8)
$$
$
h=(h_0, h_1,...,h_{m-1}), i=0,1,2,...,m-1.
$

Namely, for any collection of functions satisfying the functional
equation (6) there exists a unique splitting Gibbs measure, the
correspondence being one-to-one.

\textbf{Definition 2.} Let $K$ be a subgroup of $G_{k}$. A
collection of vectors $h=\{h_{x}\in R^m :x\in G_k\}$ is said to be
$K-periodic$ , if $h_{yx}=h_{x}$ for all $x\in G_{k}$ and $y\in
K$. A $G_k-$ periodic collections is said to be
\textit{translation-invariant}.

\textbf{Definition 3.} A Gibbs measure is called $K-periodic$ (\textit{translation-invariant}), if
it corresponds to $K- periodic$ (\textit{translation-invariant}) collection $h$.

\textbf{Proposition 1.} A SGM $\mu$ is traslation-invariant if
only if $h_{j,x}$ does not depend on $x: h_{j,x}\equiv h_i, x\in
V, j\in \Phi.$

\textbf{Proof.} Straightforward.

\textbf{Proposition 2.} Any extreme Gibbs measure is an SGM.

\textbf{Proof.} See [4], Theorem 12.6.

In [15] traslation-invariant splitting Gibbs measures are investigated.

\textbf{Periodic SGMs.} In this section we study periodic
solutions of the functional equation (6), i.e. periodic splitting
Gibbs measures. We give a description of periodic SGMs, i.e. a
characterization of such measures with respect to any normal
subgroup of finite index of $G_{k}$.

For convenience of the reader we recall some necessary notations:
Let $K$ be a subgroup of index $r$ in $G_{k}$, and let
$G_k/K=\{K_0,K_1,...,K_{r-1}\}$ be the quotient group, with the
coset $K_0=K$. Let $q_i(x)=|S_1(x)\cap K_i|, \ \ i=0,1,...,r-1; \
\ N(x)=|\{j:q_j(x)\neq0\}|$, where $S_1(x)=\{y\in G_k:<x,y>\},\ \
x\in G_k$ and $|\cdot|$ is the number of elements in the set.
Denote $Q(x)=(q_0(x),q_1(x),...,q_{r-1}(x))$.

We note (see [24]) that for every $x\in G_{k}$ there is a permutation $\pi_{x}$ of the
coordinates of the vector $Q(e)$ (where $e$ is the identity of
$G_{k}$) such that
$$
\pi_{x}Q(e)=Q(x). \eqno(9)
$$
Each $K-$periodic collection is given by $ \{h_{x}=h_{i}$  for
$x\in K_{i}, i=0,1,...,r-1\}.
$

By Theorem 1 (for m=2) and (9), vector $h_{n}, n=0,1,...,r-1$,
satisfies the system
$$
h_{n}=\sum_{j=1}^{N(e)}
q_{{i}_{j}}(e)F(h_{\pi_{n}(i_{j})};\theta,r)-F(h_{\pi_{n}(i_{j_{0}})};\theta,r),
\eqno(10)
$$
where $j_{0}=1,...,N(e)$, and function $h\mapsto F(h,m,\theta,r)$
defined in Theorem 1 now takes the form $h\mapsto
F(h)=(F_{0}(h,\theta,r),F_{1}(h,\theta,r))$ where
$$
F_{0}(h,\theta,r)=\ln\frac{r e^{h_{0}}+\theta
e^{h_{1}}+\theta^{2}}{\theta^{2} e^{h_{0}}+\theta e^{h_{1}}+r},
\eqno(11)$$
$$
F_{1}(h,\theta,r)=\ln\frac{\theta e^{h_{0}}+r
e^{h_{1}}+\theta}{\theta^{2} e^{h_{0}}+\theta e^{h_{1}}+r}. $$

\textbf{Proposition 3.} If $\theta\neq1$, then $F(h)=F(l)$ if and
only if $h=l$.

\textbf{Proof.} \emph{Necessity.} From $F(h)=F(l)$ we get the
system of equations
$$
\left\{%
\begin{array}{ll}
   $$
\frac{r e^{h_{0}}+\theta e^{h_{1}}+\theta^{2}}{\theta^{2}
e^{h_{0}}+\theta e^{h_{1}}+r}=\frac{r e^{l_{0}}+\theta
e^{l_{1}}+\theta^{2}}{\theta^{2} e^{l_{0}}+\theta e^{l_{1}}+r},
$$\\
\\
$$
\frac{\theta e^{h_{0}}+r e^{h_{1}}+\theta}{\theta^{2}
e^{h_{0}}+\theta e^{h_{1}}+r}=\frac{\theta e^{l_{0}}+r
e^{l_{1}}+\theta}{\theta^{2} e^{l_{0}}+\theta e^{l_{1}}+r},
$$\\
\end{array}%
\right.\eqno(12) $$ where $h=(h_{0},h_{1}), l=(l_{0},l_{1})$. We
obtain
\\
$$
\left\{%
\begin{array}{ll}
(r-\theta^2)\big(\theta(e^{h_0+l_1}-e^{h_1+l_0})+\theta(e^{h_1}-e^{l_1})+(r+\theta^2)(e^{h_0}-e^{l_0})\big)=0,\\
\theta^2(r-1)(e^{h_1+l_0}-e^{h_0+l_1})+\theta(r-\theta^2)(e^{h_0}-e^{l_0})+(r^2-\theta^2)(e^{h_1}-e^{l_1})=0.\\
\end{array}%
\right. \eqno(13)$$

Using the fact that
$$e^{h_0+l_1}-e^{h_1+l_0}=e^{l_1}(e^{h_0}-e^{l_0})-e^{l_0}(e^{h_1}-e^{l_1}),$$
we will have the following system of equations

$$
\left\{%
\begin{array}{ll}
(r-\theta^2)(\theta e^{l_1}+\theta^2+r)(e^{h_0}-e^{l_0})+\theta(1-e^{l_0})(e^{h_1}-e^{l_1})=0,\\
\theta(r-\theta^2-\theta e^{l_1}(r-1))(e^{h_0}-e^{l_0})+(r^2-\theta^2+e^{l_0}\theta^2(r-1))(e^{h_1}-e^{l_1})=0.\\
\end{array}%
\right. \eqno(14)$$

It follows from (14), for the case $\theta\neq 1$, that $h_0=l_0,
h_1=l_1.$

 \emph{Sufficiency.} Straightforward.
\\
Let $G_{k}^{(2)}$ be the subgroup in $G_{k}$ consisting of all
words of even length, i.e. $G_k^{(2)}=\{x \in G_k : \mbox{the
length of word $x$ is even}\}$. Clearly, $G_{k}^{(2)}$ is a
subgroup of index 2.

\textbf{Theorem 2.} \emph {Let $K$ be a normal subgroup of finite
index in $G_{k}$. Then each $K-$ periodic GM for Potts-SOS model
is either TI or $G_{k}^{(2)}-$periodic.}

\textbf{Proof.}We see from $(10)$ that
\\
$$F(h_{\pi_{n}(i_{1})})=F(h_{\pi_{n}(i_{2})})=...=F(h_{\pi_{n}(i_{N(e)})}).$$
\\
Hence from Proposition 1 we have
$$h_{\pi_{n}(i_{1})}=h_{\pi_{n}(i_{2})}=...=h_{\pi_{n}(i_{N(e)})}.$$
Therefore,

$$
h_{x}=\left\{%
\begin{array}{ll}
    h_{y}=h, & if \ {x,y \in S_{1}(z)}, \ z \in K; \\
    h_{y}=l, & if \ {x,y \in S_{1}(z)}, \ z \in G_{k}\setminus K. \\
   \end{array}%
\right. $$

Thus the measures are TI (if $h=l$) or $G_{k}^{(2)}-$periodic (if
$h\neq l$). This completes the proof of Theorem 2.

Let $K$ be a normal subgroup of finite index in $G_{k}$. What
condition on $K$ will guarantee that each $K-$ periodic GM is TI?
We put $I(K)=K\cap\{a_1,...,a_{k+1}\},$ where $a_i, \ i=1,...,k+1$
are generators of $G_{k}$.

\textbf{Theorem 3.} \emph {If $I(K)\neq\emptyset$, then each
$K-$periodic GM for Potts-SOS model is TI.}

\textbf{Proof.} Take $x\in K$. We note that the inclusion $xa_i\in
K$ holds if and only if $a_i\in K$. Since $I(K)\neq\emptyset$,
there is an element $a_i\in K$.Therefore $K$ contains the subset
$Ka_i=\{xa_i:x\in K\}$. By Theorem 2 we have $h_x=h$ and
$h_{xa_i}=l$. Since $x$ and $xa_i$ belong to $K$, it follows that
$h_x=h_{xa_i}=h=l.$ Thus each $K-$periodic GM is TI. This proves
Theorem 3.

Theorems 2 and 3 reduce the problem of describing $K-$periodic GM
with $I(K)\neq\emptyset$ to describing the fixed points of
$kF(h,\theta,r)$ which describs TIGM. If $I(K)=\emptyset$, this
problem is reduced to describing the solutions of the system:
$$
\left\{%
\begin{array}{ll}
    h=kF(l,\theta,r),\\[3mm]
    l=kF(h,\theta,r).\\[3mm]
   \end{array}%
\right.\eqno (15) $$

Denote $z_i=e^{h_i}, t_i=e^{l_i}, i=0,1$. Then from (15) we get
$$
\left\{%
\begin{array}{ll}
    z_{0}=\big(\frac{rt_0+\theta t_1+\theta^2}{\theta^2t_0+\theta t_1+r}\big)^k,
    \\[3mm]
    z_{1}=\big(\frac{\theta t_0+rt_1+\theta}{\theta^2t_0+\theta t_1+r}\big)^k, \\[3mm]
    t_{0}=\big(\frac{rz_0+\theta z_1+\theta^2}{\theta^2z_0+\theta z_1+r}\big)^k, \\[3mm]
    t_{1}=\big(\frac{\theta z_0+rz_1+\theta}{\theta^2z_0+\theta z_1+r}\big)^k. \\[3mm]
   \end{array}%
\right. \eqno (16)$$

From the first and third equations of (16), we get the following:
$$
\left\{%
\begin{array}{ll}
    z_{0}^{\frac{1}{k}}-1=\frac{(z_0-1)(r-\theta^2)}{\theta^2t_0+\theta t_1+r}, \\[3mm]
    t_{0}^{\frac{1}{k}}-1=\frac{(t_0-1)(r-\theta^2)}{\theta^2z_0+\theta z_1+r}. \\   \end{array}%
\right. \eqno (17)$$

From this, we will see that $(z_0;t_0)=(1;1)$ is a solution of the
system of equations (17) for every $\theta,r,z_1,t_1$. In this
case, we will have the following system of equations from the
second and fourth equations system of equations  (16):

$$
\left\{%
\begin{array}{ll}
    z_{1}=\big(\frac{2\theta+rt_1}{\theta^2+\theta t_1+r}\big)^k,
    \\[3mm]
    t_{1}=\big(\frac{2\theta+rz_1}{\theta^2+\theta z_1+r}\big)^k. \\
   \end{array}%
\right. \eqno (18)$$

Denote:
$$ f{(z_1)}=\big(\frac{2\theta+rz_1}{\theta^2+\theta
z_1+r}\big)^k.$$

Then the system of equations (18) will come to the following
equation:
$$ f\big(f(z_1)\big)-z_1=0.\eqno(19)$$

Obviously, the solutions of $f(z_1)=z_1$ are also the solutions of
the equation (19). We are interested in the solutions of the
equation (19), except for the solutions of the equation
$f(z_1)=z_1$. This solution is complied with $G_k^{(2)}-$periodic
measure Gibbs, which is not translation-invariant. We should
exclude the solutions of $f(z_1)=z_1$ from those of equation (19)
for the case $k=2$, simplify it, and get the following quadratic
equation:
$$(\theta^6+2\theta^4r+\theta^2r^2+r^4+2\theta
r^3+2\theta^3r^2)z_1^2+$$
$$+(2\theta^7+6\theta^5r+6\theta^3r^2+6\theta
r^3-4\theta^4+\theta^4r^2+8\theta^2r^2+2\theta^2r^3+8\theta^4r+r^4)z_1+$$
$$+4\theta^2r^2+4\theta^6r+r^4+6\theta^4r^2+4\theta^2r^3+\theta^8+4\theta^5r+8\theta^3r^2+4\theta r^3=0.$$

In order to have two positive real roots in that equation, there
should the following conditions: $D>0, b<0$, where
$$D=(2\theta^7+6\theta^5r+6\theta^3r^2+6\theta
r^3-4\theta^4+\theta^4r^2+8\theta^2r^2+2\theta^2r^3+8\theta^4r+r^4)^2-$$
$$-(\theta^6+2\theta^4r+\theta^2r^2+r^4+2\theta
r^3+2\theta^3r^2)\cdot(4\theta^2r^2+4\theta^6r+r^4+6\theta^4r^2+4\theta^2r^3+\theta^8+4\theta^5r+8\theta^3r^2+4\theta
r^3),$$
$$b=2\theta^7+6\theta^5r+6\theta^3r^2+6\theta
r^3-4\theta^4+\theta^4r^2+8\theta^2r^2+2\theta^2r^3+8\theta^4r+r^4.$$

As a result, we obtain the following theorem

\textbf{Theorem 4.} \emph {Let $k=2$. If $D>0, b<0$, then there
exist at least two $G_k^{(2)}-$periodic (not
translation-invariant) Gibbs measures for Potts-SOS model; If
$D=0, b<0$, then there exists at least one $G_k^{(2)}-$periodic
(not translation-invariant) Gibbs measure for Potts-SOS model.}

Let's show that the set  $\{{(r,\theta)\in R^2:D \geq 0,b<0}\}$ is
not empty. Indeed, let $r=\theta^2$.
 Then we have the following:
$$D=-16\theta^8(\theta^2-1)^2(3\theta^4+10\theta^3+6\theta^2-1),$$
$$b=4\theta^4(\theta^4+5\theta^3+4\theta^2-1).$$

If $\theta<\theta_D (\theta_D\approx0,32359)$, then we have $D>0,
b<0$, that is to say that there exist at least two
$G_k^{(2)}-$periodic (not translation-invariant) Gibbs measures;
If $\theta=\theta_D$, in that case, the followings are obvious:
$D=0, b<0$, which means there exists at least one
$G_k^{(2)}-$periodic (not translation-invariant) Gibbs measure for
Potts-SOS model.

\textbf{Remark.} Note that, when $k=2$ for the Potts model there
is no periodic Gibbs measure (see [19]), but as Theorem 4 shows
for the Potts-SOS model, there exists such kind of measures under
some conditions.

\textbf{Aknowledgments.} The author is grateful to Professors
U.A.Rozikov and M.M.Rahmatullaev for useful discussions and for
valuable comments.

{\bf References}

[1] Ya. G. Sinai, \textit{Theory of Phase Transitions: Rigorous
Results,} (Pergamon, 1982).

[2] C. Preston, \textit{Gibbs States on Countable Sets,} Cambridge
Univ.Press, 1974;

[3] V. A. Malyshev and R. A. Minlos, \textit{Gibbs Random Fields,}
Nauka, 1985.

[4] H. O. Georgii, \textit{Gibbs Measures and Phase Transitions, }
Walter de Gruyter, 1988.

[5] S. Zachary, \textit{Ann. Probab.} \textbf{11}:4 (1983),
894--903.

[6] S. Zachary, \textit{ Stoch. Process. Appl.} \textbf{20}:2
(1985), 247--256.

[7] P. M. Bleher and N. N. Ganikhodjaev, \textit{Theor. Probab.
Appl.} \textbf{35} (1990), 216--227.

[8] P. M. Bleher,  \textit{Commun. Math. Phys.} \textbf{128}
(1990), 411--419.

[9] P. M. Bleher, J. Ruiz and V. A. Zagrebnov,
\textit{J.Stat.Phys.}, \textbf{79} (1995), 473--482.

[10] P. M. Bleher, J. Ruiz and V. A. Zagrebnov,
\textit{J.Stat.Phys.} \textbf{93}, (1998), 33--78.

[11] D. Ioffe,  \textit{Lett. Math. Phys.} \textbf{37} (1996),
137--143.

[12] D. Ioffe, \textit{Extremality of the disordered state for the
Ising model on general trees,} in Trees, Versailles, 1995, Prog.
Probab. Vol. \textbf{40}, (Birkhauser, 1996), pp. 314.

[13] P. M. Bleher, J. Ruiz, R. H. Schonmann, S. Shlosman and V. A.
Zagrebnov,  \textit{Moscow Math. J.} \textbf{3}, (2001), 345--363.

[14] U.A. Rozikov,  \textit{Gibbs measures on Cayley trees.} World
scientific, 2013.

[15] H. Saygili,  \textit{Asian Journal of Current Research} V. 1,
N 3, (2017), 114--121.

[16] M. M. Rahmatullaev,  \textit{Russian Mathematics},
\textbf{59}:11 (2015), 45--53.

[17] M. M. Rahmatullaev, \textit{Uzb. Mat. Zh.,}
\textbf{2},(2009), 144--152

[18] C. Kulske, U. A. Rozikov, R. M. Khakimov. \textit{Jour. Stat.
Phys.}  \textbf{156}:1, (2014), 189--200.

[19] U.A.Rozikov and R.M.Khakimov,  \textit{Theor. Math.Phys.,}
\textbf{175}, (2013), 699–-709.

[20] R.M.Khakimov,  \textit{Uzb. Mat. Zh.,} \textbf{3}, (2014),
134–-142.

[21] M.M.Rahmatullaev,  \textit{Theor. Math. Phys.,} \textbf{180},
(2014), 1019–-1029.

[22] U.A.Rozikov, Y.M.Suhov, \textit{Infinite Dimensional
Analysis, Quantum Probability and Related Topics.,} Vol.9, No. 3
(2006), 471--488.

[23] C. Kulske, U. A. Rozikov, \textit{J.Stat.Phys.} \textbf{160},
(2015) 659--680.

[24] U. A. Rozikov,  \textit{Theor. Math. Phys}. \textbf{112}
(1997), 929--933.

\end{document}